\documentclass[ aps,%
floatfix,%
final,%
notitlepage,%
oneside,%
onecolumn,%
nobibnotes,%
nofootinbib,%
superscriptaddress,%
showpacs,%
]%
{revtex4}

\usepackage{epsfig}
\usepackage{graphics}

\newcommand{\JP}{\psi} \newcommand{\ec}{\eta_c}

\newcommand{\epem}{e^+e^-}

\newcommand{\GeV}{\,\mathrm{GeV}}

\newcommand{\beq}{\begin{eqnarray}}\newcommand{\eeq}{\end{eqnarray}}
\newcommand{\beqa}{\begin{eqnarray*}}\newcommand{\eeqa}{\end{eqnarray*}}

\begin{document}

\title{The processes $\epem\to J/ \Psi \chi_{c0}, \Psi(2S) \chi_{c0}$  at $\sqrt s = 10.6$GeV in the framework of light cone formalism.}
\author{V.V. Braguta}
\email{braguta@mail.ru}

\author{A.K. Likhoded}
\email{Likhoded@ihep.ru}

\author{A.V. Luchinsky}
\email{Alexey.Luchinsky@ihep.ru}
\affiliation{Institute for High Energy Physics, Protvino, Russia}

\begin{abstract}
In this paper we have analyzed the production of pair charmonium mesons in  
the reactions $\epem\to J/ \Psi \chi_{c0}, \epem\to \Psi(2S) \chi_{c0}$
at energy $\sqrt s = 10.6$GeV in the framework of the light cone formalism. In comparison 
with NRQCD the numerical results for the cross sections are in better agreement 
with experiment.
\end{abstract}


\maketitle

\newcommand{\ins}[1]{\underline{#1}}
\newcommand{\subs}[2]{\underline{#2}}
\section{Introduction}
Exclusive double charmonium meson production in $e^+ e^-$ annihilation at energy $\sqrt s = 10.6$ GeV remains 
very interesting problem for theoretical investigations. NRQCD that is often used to predict the cross section 
of such processes fails to archive agreement with Belle and BaBar experiments. The predictions for the 
cross sections obtained in the framework of NRQCD\cite{Braaten:2002fi} are  by an order of magnitude less than 
the experimental results\cite{Abe:2004ww,Aubert:2005tj}. 

An interesting approach to the problem was proposed in papers\cite{Ma:2004qf,Bondar:2004sv}. In these papers 
the process $\epem \to \JP \ec$ at energy $\sqrt s = 10.6 \GeV$ was considered in the framework of light cone formalism. 
In paper \cite{Ma:2004qf} it was shown that taking different light cone wave functions one can considerably 
enhance NRQCD prediction for the cross section of the process $\epem \to \JP \ec$. Using physical model for the light 
cone wave functions of $\JP, \ec$ mesons the authors of paper \cite{Bondar:2004sv} received the prediction for the
cross section of this process that agrees with experimental result obtained at Belle. In addition
to the agreement between the theoretical prediction and the experimental results light cone approach allows one to understand that 
charmonium mesons wave functions are two wide to be considered in the framework of NRQCD in such processes.

Further progress in the understanding of exclusive double charmonium mesons production is connected with paper\cite{Braguta:2005kr}.
In this paper it was shown that contrary to NRQCD the study of the processes $\epem \to \Psi(1S) \ec(2S), \Psi(2S) \ec(1S), \Psi(2S) \ec(2S)$
in the framework of light cone formalism leads to a good agreement with Belle and Babar results\cite{Abe:2004ww,Aubert:2005tj}. 

So light cone formalism better than NRQCD predicts  the cross sections 
$\epem \to J/ \Psi \ec(1S), J/ \Psi \ec(2S), \Psi(2S) \ec(1S), \Psi(2S) \ec(2S)$. 
In addition to these reactions Belle and BaBar have measured the cross section 
of the processes $\epem \to J /\Psi \chi_{c0}, \Psi(2S) \chi_{c0}$. The agreement of NRQCD 
prediction and experimental results is also poor for this reactions. The aim of this paper 
is the application of light cone formalism to these reactions.

This paper is organized as follows. In the next section light cone wave function of $J/ \Psi$, 
$\Psi(2S)$ and $\chi_{c0}$ mesons will be considered. In the third section the expression 
for the amplitudes of the processes under consideration will be derived. Numerical analysis 
of the cross sections $\epem \to J /\Psi \chi_{c0}, \Psi(2S) \chi_{c0}$ will be presented 
in the forth section. Finally the results of the calculation will be discussed in last section.

\section{Light cone wave functions of $J/ \Psi, \Psi(2S)$ and $\chi_{c0}$ mesons.}
To calculate the amplitude of the process $\epem \to V \chi_{c0}$[$V=J/ \Psi, \Psi(2S)$]
one needs to know light cone wave functions of the final charmonium mesons. The functions 
for $J/ \Psi$ and $\Psi(2S)$ mesons are defined as follows\cite{Bondar:2004sv}
\beq
{\langle V_{\lambda}(p)|{\bar Q}_{\beta}(z)\,Q_{\alpha}(-z)|0\rangle}_{\mu}=
\frac{f_{V} M_V}{4}\int^1_o dx_1\,e^{i(pz)(x_1-x_2)}
\biggl \{ {\widehat p}\,\frac{(e_{\lambda}z)}{(pz)}\, V_L(x) + \bigl ( {\widehat e}_{\lambda} - {\widehat p}\,\frac{(e_{\lambda}z)}{(pz)} \bigr ) \, V_{\perp}(x)+\nonumber \\
 \frac {f^{t}_{v}(\mu)} {M_V} (\sigma_{\mu\nu} e_{\lambda}^{\mu}\, p^{\nu})\,V_{T}(x)+  
f^{a}_{v}(\mu) (\epsilon_{\mu\nu\alpha\beta}\gamma_{\mu}\gamma_5\,e_{\lambda}
^{\nu}\,p^{\alpha}z^{\beta})\, V_{A}(x) \biggl \}_{\alpha\beta}.
\label{psi}
\eeq
The dependence of the light cone wave functions on the scale $\mu$ is very slow and 
it will not be considered in the full form. Only renormalization factors of the corresponding local currents
will be regarded. The constants $f^t_v, f^a_v$ can be determined from QCD equations of motion
\beq
f^t_v(\mu) = \frac {2  {\overline M_Q} } {M_{\Psi}} Z_t, \nonumber \\
f^a_v(\mu) = \frac 1 2 \bigl( 1- Z_t Z_m \frac {4 {\overline M_Q}^2} {M_{\Psi}^2} \bigr ),
\eeq
where $\overline M_Q = M^{\overline {MS} }_Q ( \mu = M^{\overline {MS} }_Q)$. The factors $Z_p, Z_t, Z_m$ can be written in the form
\beq
Z_p  &=& \left[\frac{\alpha_s(\mu^2)}{\alpha_s({\overline M}_Q^2)} \right]^{\frac{-3C_F}{b_o}},\quad ~~ 
Z_{t} = \left[\frac{\alpha_s(\mu^2)}{\alpha_s({\overline M}_Q^2)} \right]^{\frac{C_F}{b_o}},\quad 
~~ Z_{m} = \left[\frac{\alpha_s(\mu^2)}{\alpha_s({\overline M}_Q^2)} \right]^{\frac{3C_F}{b_o}},
\eeq
where $C_F=4/3$, $b_o=25/3$.

For the light cone wave functions of $J/ \Psi$ meson the model proposed in \cite{Bondar:2004sv} will be used:
\beq
\phi_{i}(x,v^2) &=& 
  c_{i}(v^2)\,\phi^{a}_{i}(x)\left\{
    \frac{x_1 x_2}{[1-4x_1x_2(1-v^2)]} 
  \right\}^{1-v^2}\,
  \label{psi1}.
\eeq
For the light cone wave functions of $\Psi(2S)$ meson the model proposed in \cite{Braguta:2005kr} will be used:
\beq
\phi_{i}(x,v^2)=c_{i}(v^2)\,\phi^{a}_{i}(x)\, \Biggl ( 1 - 8  v^2 \beta \frac {(1-v^2) x_1 x_2} {[1-4x_1x_2(1-v^2)]} \Biggr) \Biggl \{ \frac{x_1 x_2}
{[1-4x_1x_2(1-v^2)]} \Biggr \}^{1-v^2}.
\label{psi2}
\eeq
Here $v$ is a characteristic speed of quark-antiquark pair in meson, $c_{i}$ is 
the coefficient fixed by the wave function normalization condition $\int d x \phi_{i}(x,v^2) =1$.
The constant $\beta$ is fixed by the condition that zero of the wave function of $2S$ state meson\cite{Braguta:2005kr} must 
coincide with zero obtained from the solution of Schrodinger equation with Buchmuller-Tye potential\cite{Buchmuller:1980su}.
$\phi^{a}_i$ are the asymptotic expressions of the wave functions
\beq
V_A(x)=V_L(x)=V_T(x)=\phi_{asy}(x)=6x_1 x_2,
\eeq
\beq
\quad V_{\perp}(x)=\frac{3}{4}[1+(x_1-x_2)^2].\,
\eeq
It should be noted here that the expressions for light cone wave functions (\ref{psi1}),(\ref{psi2}) link 
different limits: quark-antiquark pair in meson being at rest $v \to 0$ and 
very light quark $v \to 1$. In the former limit one obviously gets $\sim \delta (x - 1/2)$, the later one 
leads to the asymptotic function $\sim \phi^{a}$.

Let us consider the light cone wave functions of $\chi_{c0}$ meson:
\beq
{\langle \chi_{c0}(p)|{\bar Q}(z) Q (-z)|0\rangle} = f_S \int d y e^{i pz (y_1 - y_2)} S_S(y), 
\nonumber \\
{\langle \chi_{c0}(p)|{\bar Q}(z) \gamma_{\mu} Q (-z)|0 \rangle} = 
f_V^{(1)} p_{\mu} \int d y e^{i pz (y_1 - y_2)} S_V (y) + f_V^{(2)} z_{\mu} \int d y e^{i pz (y_1 - y_2)} S_V^{(2)}(y),
\nonumber \\
{\langle \chi_{c0}(p)|{\bar Q}(z) \sigma_{\mu \nu} Q (-z)|0\rangle} = f_T (p_{\mu} z_{\nu} - p_{\nu} z_{\mu}) \int d y e^{i pz (y_1 - y_2)} S_T(y).
\eeq
The functions $\phi_i = S_S(y), S_V^{(2)}(y), S_T(y)$ are normalized as follow $\int dy \phi_i =1$,
the normalization condition for $S_V(y)$ is $\int dy (y_1-y_2) S_V(y)=1$.
Using QCD equation of motion the constants $f_S, f_T, f_V^{(2)}$ can be related to $f_V^{(1)}$:
\beq
f_S = -3 \frac {f_V^{(1)} M_{\chi}^2} {2 \overline M_Q}, \nonumber \\
f_V^{(2)} = -i M_{\chi}^2 f_V^{(1)}, \nonumber \\
f_T = i \frac {f_V^{(1)} M_{\chi}^2} {2 \overline M_Q}. 
\label{ct}
\eeq
The constant $f_S$ and consequently $f_V^{(1)}$ can be expressed through matrix element 
$\langle \chi_{c0}(p)|{\bar Q}(0) Q (0)|0\rangle$ found in \cite{Novikov:1977dq}.
As the result we get
\beq
f_V^{(1)} = \sqrt \frac { 3 R'(0)^2 } {2 \pi m_Q^3} = \sqrt \frac { <O_1>_P } {3 m_Q^3},
\label{fv}
\eeq
where $R'(0)$ is derivative of the wave function of heavy quarkonium at the origin, $m_Q$
is the mass of $c$ quark in potential model, $<O_1>_P$ is well known from NRQCD matrix element
\cite{Bodwin:1994jh}. It should be 
noted that formulae (\ref{ct}) were found in paper\cite{Chernyak:1983ej} but 
the expression for the constant $f_V^{(1)}$ differs from (\ref{fv}) and agrees with \cite{Chernyak:new}.
 
 Although there are four light cone $\chi_{c0}$ meson wave functions only $S_S(y)$ and 
$S_V(y)$ contribute to the amplitude of the processes at accuracy considered in our paper, 
the others give power correction to the result. Below we will consider only $S_S(y)$ and $S_V(y)$. 

 As in the case of vector mesons we are not going to regard full $\mu$-dependence of light cone wave
functions. Only overall renormalization factor of corresponding operator will be considered. 
The renormalization factor for light cone wave function $S_S(y)$ equals to $Z_p(\mu)$. To find 
renormalization factor for the function $S_V(y)$ it is worth noting that the corresponding 
operator ${\langle \chi_{c0}(p)|{\bar Q}(0) \gamma_{\mu} Q (0)|0 \rangle}$ equals zero. 
So the renormalization factor for the function $S_V(y)$ equals  the renormalization factor 
of the first nonvanishing multiplicatively renormalized local operator 
${\langle \chi_{c0}(p)|{\bar Q} \gamma_{\mu} C_1^{3/2}(z D / z \partial) Q |0 \rangle}$. 
It equals to\cite{Chernyak:1983ej}
\beq
Z_v =  \left[\frac{\alpha_s(\mu^2)}{\alpha_s({\overline M}_Q^2)} \right]^{\frac{ 8 C_F}{9 b_o}}.
\eeq

In order to get the expressions for the light cone wave functions $S_V(y)$ and $S_S(y)$ 
we will use the same procedure that was used in paper \cite{Bondar:2004sv} to get 
expressions (\ref{psi1}). 

First let us consider the leading twist wave function of $\chi_{c0}$ meson $S_V(y)$. There are 
three wave functions of $2P$ meson with different orbital momentum projection $L_z$ in the direction 
of  meson motion($m=\pm 1, 0$). But only $m=0$ projection gives contribution to the leading 
twist light cone wave function\cite{Chernyak:1983ej}. This wave function  can be approximated by 
Coulomb wave function of $2P$ state:
\beq
\Psi_c \sim \cos { \theta} \frac k {(k^2+m_c^2 v^2)^3},
\eeq
where $v$ is the characteristic relative velocity. Substituting\cite{terentev}
\beq
\vec {k}_{\perp} \to \vec {k}_{\perp}, \quad k_z \to (y_1 - y_2) \frac {M_0} 2, \quad M_0^2 = \frac {m_c^2 + \vec {k}_{\perp}^2 } {y_1 y_2}
\eeq
and than carrying out the integration
\beq
\phi \sim \int d^2 k_{\perp} \Psi_c(y,\vec {k}_{\perp})
\eeq
one gets the expression for the light cone wave function:
\beq
\phi \sim S^{as}_V \biggl \{ y_1 y_2 \frac {1- 2 y_1 y_2( 1- v^2) } {(1- 4 y_1 y_2( 1- v^2))^2}
\biggr \}.
\label{P1}
\eeq
In last formula the expression $S_V^{as} = y_1 y_2 (y_1 - y_2)$ is asymptotic form of the light cone wave function
$S_V(y)$. To get final expression for the wave function we will modify (\ref{P1}) similar to wave functions of $1S$ 
and $2S$ mesons (\ref{psi1}), (\ref{psi2}):
\beq
S_V(y) = c_V(v^2) S_V^{as}(y) \biggl \{ y_1 y_2 \frac {1- 2 y_1 y_2( 1- v^2) } {(1- 4 y_1 y_2( 1- v^2))^2}
\biggr \}^{1-v^2},
\label{chiv}
\eeq
where $c_V (v^2)$ is the coefficient which is fixed by the wave function normalization $\int d y S_V(y) (y_1- y_2) =1$.  
For the $S_S(y)$ wave function the following expression will be taken 
\beq
S_S(y) = c_S(v^2) S_S^{as}(y) \biggl \{ y_1 y_2 \frac {1- 2 y_1 y_2( 1- v^2) } {(1- 4 y_1 y_2( 1- v^2))^2}
\biggr \}^{1-v^2},
\label{chis}
\eeq
where asymptotic form of the wave function $S_S^{as}(y)=1$, $c_S (v^2)$ is the coefficient which is fixed by 
the wave function normalization $\int d y S_S(y)=1$.

\section{The calculation of the amplitude of the processes $\epem \to V(p_1, \epsilon_1 ) \chi_{c0}(p_2)$.}
Leading  asymptotic behavior of the matrix element $\left<V(p_1, \epsilon_1), \chi_{c0}(p_2) | J_{\mu}^{el} | 0\right>$ can be derived from formula\cite{Chernyak:1977as}
\beq
\left<M(p_1, \lambda_1) M(p_2, \lambda_2)|J_{\mu}^{el}|0\right> & \sim & 
  \left( \frac 1 {\sqrt s}  \right)^{|\lambda_1 - \lambda_2|+1},
\label{as}
\eeq
where $J_{\mu}^{el}$ is the electromagnetic current. For the processes under consideration we have $M(p_1, \lambda_1)=V(p_1, \lambda_1), M(p_2, \lambda_2)=\chi_{c0}(p_2)$. 
Obviously the helicity $\lambda_2$ equals zero. As 
to the vector meson $V(p_1, \lambda_1)$ the leading contribution is given by the helicity $\lambda_1=0$. So the asymptotic behavior of the amplitude is
\beq
\left<V(p_1, \lambda_1), S(p_2)| J_{\mu}^{el} | 0\right> & \sim & \frac 1 {\sqrt s}
\eeq
and the asymptotic behavior of the cross section $\sigma (\epem\to VS)$ is $\sim 1/s^3$. Unfortunately 
one can show that it is not possible to disregard  NLO contribution in $1/s$ expansion. 
To see this let us consider NRQCD result for the cross section of the process $\epem\to J/ \JP(1S) \chi_{c0}$  obtained in paper 
\cite{Braaten:2002fi}. Cross section of this process can be written as
\beq
\sigma & = & \frac {\pi^3} {3^5 s} \alpha^2 \alpha_s^2 q_c^2 F_0 r^2 \sqrt{1-r^2} \frac {f_V^2 f_S^2} {m_c^4},
\eeq
where $r^2 = 16 m_c^2 /s$ and $F_0 = 2(18 r^2 - 7 r^4)^2+ r^2(4 + 10 r^2 - 3 r^4)^2$. Let us substitute 
$s \to 10.6^2 \xi$ and expand the above formula in $1/ \xi$ series (numerical inputs are the same as in \cite{Braaten:2002fi}). We get
\beq
\sigma & = & \frac {0.23} {\xi^3} + \frac {2.91} {\xi^4} - \frac {0.89} {\xi^5} + O(1/ \xi^6). 
\label{sig}
\eeq
Thus one sees that in the framework of NRQCD NLO correction at energy $\sqrt s = 10.6$ GeV  is by an order of magnitude larger
than the leading one. So one can suppose that NLO term in $1/s$ expansion gives considerable contribution and must be 
regarded in our analysis. The same is true for the process $\epem\to \JP(2S) \chi_{c0}$.

Two diagrams that contribute to the processes $\epem \to V(p_1, \lambda_1) \chi_{c0}(p_2)$ are shown 
in Fig.1. and the other two  can be obtained from them by charge conjugation.
Having the expressions for the light cone wave functions (\ref{psi1}), (\ref{psi2}), (\ref{chiv}), (\ref{chis}) it is not difficult 
to obtain the matrix element $\left<V(p_1, \lambda_1), \chi_{c0}(p_2) | J_{\mu} | 0\right>$:
\beq
\langle V(p_1, \lambda_1) \chi_{c0}(p_2) | J^{\mu} | 0 \rangle =
g_1 \bigl ( p_1^{\mu} - p_2^{\mu} \bigr ) (e_{\lambda_1} p_2) + g_2 \bigl ( (e_{\lambda_1} p_2) p_1^{\mu} - e_{\lambda_1}^{\mu} (p_1 p_2) \bigr )
\label{Jel}
\eeq
where the formfactors $g_1$ and $g_2$ are:
\beq
g_1 = 2 \frac {\pi} 9 \frac {f_V^{(1)} f_V} {s^2}  M_V M_{\chi}  \int_0^1 d x_1 \int_0^1 d y_1  \alpha_s(k) \biggl ( - 16 Z_v(k) \frac {S_V(y) V_L(x)} {M_{\chi}} \frac 1 {d(x,y)}  \biggr ) 
\label{g1}
\eeq
\beq
\nonumber
g_2 = 2 \frac {\pi} 9 \frac {f_V^{(1)} f_V} {s^2}  M_V M_{\chi} \int_0^1 d x_1 \int_0^1 d y_1 \alpha_s(k) \biggl ( 2 Z_v(k) \frac {S_V(y) V_A(x) } {M_{\chi}} (1- Z_m(k) Z_t(k) \frac {4 {\overline M}_Q^2 } {M_V^2} ) \frac {1+y_1} {d(x,y)^2} + \\
16 Z_t(k) Z_v(k) Z_m(k) \frac { S_V(y) V_T(x) } {s(x) d(x,y)} \frac { {\overline M}_Q^2} { M_V^2 M_{\chi}}  
- 8 Z_v(k) \frac {S_V(y)} {M_{\chi} d(x,y)}  ( V_L(x) - 2 V_{\perp}(x) - \frac  {V_{\perp}(x) } {s(y)} )  + \nonumber \\
32 Z_p(k) Z_t(k) \frac {S_S(y) V_T(x)} {s(x) d(x,y)} \frac { {\overline M}_Q } {M_V^2} f
\biggr ),
\label{g2}
\eeq
where $d(x,y), s(x), s(y)$ are dimensionless quark and gluon propagators defined as follows:
\beq
d(x,y)&=& 
  \frac{k^2}{q_0^2}=\left( x_1+\frac{\delta}{y_1}\right)
  \left(y_1+\frac{\delta}{x_1}\right),
\qquad \delta=\frac {\Bigl (Z_m(k) {\overline  M}_Q \Bigr )^2} {s}\,, 
\\ 
s(x) & = & \left(x_1+\frac{(Z_m({\sigma}){\overline  M}_Q)^2}{y_1y_2\,s} \right),
\quad s(y)= \left(y_1+\frac{(Z_m({\sigma}){\overline  M}_Q)^2}{x_1x_2\,s}
\right), 
\eeq
where $k$ is a momentum of virtual gluon,
$\sigma$ is a characteristic momentum of virtual quark in the diagrams shown 
in Fig.1, the constant $f=-3 M_{\chi_{c0}} / 2 {\overline M}_{Q}$. 

It is interesting to note that there are two contributions to the  formfactors $g_1$ and $g_2$ 
factored in formulae (\ref{g1}), (\ref{g2}).
The first is NRQCD result for the formfactors that does not regard internal motion 
of quark-antiquark pair inside mesons and it is proportional to  $\sim f_V^{(1)} f_V \sim R(0) R'(0)$.
The second contribution regards internal motion and it is proportional to the integrals $\int d x_1 d y_1$
in expressions (\ref{g1}), (\ref{g2}). 

In the limit $v \to 0$ the mesons $V$ and $\chi_{c0}$ have equal masses $M=M_V=M_{\chi}$ and formulae (\ref {g1}), (\ref {g2})
must reproduce NRQCD result for the process $\epem \to J/ \Psi \chi_{c0}$:
\beq
\nonumber
g_1 = \frac {128 \pi} 9 \frac {f_V^{(1)} f_V} {s^2}  M  ( 1  - 4 \frac {M^2} s ), \\
g_2 = - \frac {128 \pi} 9 \frac {f_V^{(1)} f_V} {s^2}  M  ( 9  - 14 \frac {M^2} s ).
\label{nrqcd}
\eeq
Really if one takes the limit $v \to 0$ in expressions (\ref {g1}), (\ref {g2}) one gets
\beq
\nonumber
g_1 = \frac {128 \pi} 9 \frac {f_V^{(1)} f_V} {s^2}  M \biggl ( 1  + O \biggl ( \frac {M^2} s \biggr ) \biggr ), \\
g_2 = - \frac {128 \pi} 9 \frac {f_V^{(1)} f_V} {s^2}  M \biggl ( 9  + O \biggl ( \frac {M^2} s \biggr ) \biggr) .
\label{lightcone}
\eeq
So one sees that at accuracy considered in our paper the expressions (\ref{nrqcd}) and (\ref{lightcone}) coincide.

The cross section of the processes $\epem \to V \chi_{c0}$ is given by the formula:
\beq
\sigma = \frac {\pi \alpha^2_{el}} { s^3} q_c^2 \biggl ( \frac { |{\bf p_1}|} {\sqrt{s}} \biggr ) F,
\label{cr}
\eeq
where $q_c$ is the charge of $c$-quark, $\bf p_1$ is the momentum of vector meson in final mesons' center mass frame,
for the electromagnetic current of the form (\ref{Jel}) the function $F$ is given by the formula
\beq
F= \frac {g_1^2 q_0^8} {6 M_V^2} + \frac {g_2^2 
q_0^6} 3 - \frac 1 3 g_1 g_2 q_0^6 + \frac 1 2 g_2^2 M_V^2 q_0^4 - 
\frac 4 3 g_1^2 M_{\chi}^2 q_0^4 + \frac 1 3 g_2^2 M_{\chi}^2 q_0^4 - 
\frac 2 3 g_1 g_2 M_{\chi}^2 q_0^4 + \nonumber \\ \frac 2 3 g_2^2 M_V^2 M_{\chi}^2 q_0^2 +   
\frac 4 3 g_1 g_2 M_V^2 M_{\chi}^2 q_0^2 + \frac 8 3 g_1^2 M_V^2 M_{\chi}^4 + 
\frac 2 3 g_2^2 M_V^2 M_{\chi}^4 + \frac 8 3 g_1 g_2 M_V^2 M_{\chi}^4 \nonumber \\
 = \frac {g_1^2 s^4} {6 M_V^2} - \frac {2 g_1^2 s^3} 3 +  \frac {g_2^2 s^3} 3 - 
\frac { 2 g_1^2 M_{\chi}^2 s^3} {3 M_V^2} - \frac 1 3 g_1 g_2 s^3 +O(1/s^2),
\label{cr1}
\eeq
where $q_0^2 = s - M_V^2 - M_{\chi}^2$.

\section{Numerical calculation.}

In the numerical analysis the following parameters will be used:
\beq
\nonumber
\overline {M}_c = 1.2 \mbox{GeV}, \\ \nonumber
\JP(1S), ~~  f_V = 0.41 \mbox{GeV},  \\ 
\JP(2S), ~~  f_V = 0.28 \mbox{GeV}.
\label{const}
\eeq
The values $f_V$ were obtained from decay width 
\beq
\Gamma(V \to e^+ e^-) = \frac {16 \pi \alpha^2} {27} \frac {|f_V|^2} {M_V}.
\eeq
As to the value of the constant $f_V^{(1)}$ it is related to the value of the constant 
$f_S = \langle \chi_{c0}(p)|{\bar Q}(0) Q (0)|0\rangle$ found in \cite{Novikov:1977dq} using QCD sum rules:
\beq
f_V^{(1)} = 0.084. \mbox{GeV}
\eeq
For $\alpha_s(\mu)$ one loop result will be used
\beq
\alpha_s(\mu)=\frac {4 \pi} {b_0 \log (\mu^2/ \Lambda^2)},
\eeq
with $\Lambda = 200$MeV. The other parameters needed for the numerical analysis is the width of wave function $v^2$.
It will be taken from potential model\cite{Buchmuller:1980su}: 
\beq
\nonumber
J / \Psi ~~ v^2=0.23 , \\ \nonumber
\chi_{c0} ~~ v^2=0.25, \\
\JP(2S) ~~ v^2=0.29. 
\label{width}
\eeq
Now let us consider the formula of the cross section (\ref{cr}). It is seen 
from (\ref{cr1}) that leading order contribution(LO) to the cross section is given by the first term $\sim g_1^2$
and the rest of the formula (\ref{cr1}) is the power correction to the LO. 
In our paper the power correction to the  formfactor $g_1$ is beyond the accuracy of our calculation. 
But  it is well seen from (\ref{cr1}) that in order to get the cross section up 
to $O( 1/ s^5)$ one must know $1/s$ correction to the formfactor $g_1$.
Fortunately LO term $\sim g_1^2$ gives negligible contribution to the cross section
and varying $g_1^2$ in a reasonable region the cross section is changed by few percent.
Sure we do not pretend to the accuracy about few percent in our calculation. 

\begin{table}
$$\begin{array}{|c|c|c|c|c|}
\hline
   H_1 H_2 & \sigma_{BaBar} \times Br_{H_2 \to charged >2}(\mbox{fb}) \mbox{\cite{Aubert:2005tj}}  
 & \sigma_{Belle}\times Br_{H_2 \to charged >2}(\mbox{fb}) \mbox{\cite{Abe:2004ww}}
 & \sigma_{LO} (\mbox{fb})  
 & \sigma_{NRQCD} (\mbox{fb}) \mbox{\cite{Braaten:2002fi}} \\
\hline
\JP(1S) \chi_{c0}& 10.3 \pm 2.5^{+1.4}_{-1.8} & 6.4 \pm 1.7  \pm 1.0 & 14.4_{~~13.3}^{~15.5}  &2.3\\
\hline
\JP(2S) \chi_{c0}&  - & 12.5 \pm 3.8 \pm 3.1 & 7.8_{~7.3}^{~8.3} & 1.0\\
\hline
\end{array}$$
\caption{The second and third column contain experimental result. The results of our calculation 
is presented in the forth column. The last column contains NRQCD results.}
\end{table}

In our paper we use the model for the light cone wave function defined by equations (\ref{psi1}), (\ref{psi2}),
(\ref{chiv}) and (\ref{chis}) with the widths (\ref{width}). To estimate the size of this uncertainty 
results from this model in addition to the widths (\ref{width}) the calculation 
of the cross sections with shifted widths(about 10\%) is done. The results of the 
calculation is presented in Table 1.
 The second and the third columns 
contain experimental results measured at BaBar and Belle experiments.
In the fourth column the results of this paper are presented. 
Central values of the cross sections correspond to the light cone wave functions
with widths (\ref{width}). The upper values of the cross section correspond
to the widths $0.26(J/ \Psi), 0.32 ( \Psi(2S)), 0.28 ( \chi_{c0})$.
The lower values of the cross section correspond
to the widths $0.2(J/ \Psi), 0.26 (\Psi(2S)), 0.22 (\chi_{c0})$. In order to compare the 
result with NRQCD predictions for the processes under consideration the fifth 
column contains the predictions in the framework of this model. 

From Table 1 one sees that the predictions of the cross section of the processes 
$\epem\to J/ \Psi \chi_{c0}, \Psi(2S) \chi_{c0}$ in the framework of light cone 
are larger than NRQCD predictions and the agreement with the experimental results is better. 
As it was noted in \cite{Braguta:2005kr} the difference 
of the NRQCD and light cone prediction for the cross sections can be attributed to the fact 
that at leading approximation NRQCD does not regard the motion inside final mesons. 
So NRQCD predictions for the cross section of the processes under consideration 
are unreliable. 

In addition to the uncertainties described above and uncertainty due to the unknown 
size of radiative QCD correction one can suppose that there is very important $1/s$
correction. The size of this correction 
can be estimated from formula (\ref{sig}) for NRQCD result of the cross section
of the process $\epem\to J/ \Psi \chi_{c0}$. 
It is seen from (\ref{sig}) that in the framework of NRQCD $O(1/s^5)$ contribution 
changes the value of the cross section by 30 \%. Moreover $O(1/s^5)$ correction 
diminish the value of the cross section. It was noted above that to get 
light cone result one must multiply NRQCD by a factor regarding internal mesons' 
motion. If one suppose further that these factors for $O(1/s^4)$ and $O(1/s^5)$
contribution are of the same order than one can claim that in the framework of 
light cone formalism $O(1/s^5)$ contribution diminish the size of the cross section by 30\%.
After including this correction the value of the cross sections 
can be estimated as $\sigma( \epem\to J/ \Psi \chi_{c0}) \sim 9$fb and 
$\sigma(\epem\to \Psi(2S) \chi_{c0}) \sim 5$fb.

It should be noted here that light cone prediction for the cross section 
of the process $\epem\to \Psi(2S) \chi_{c0}$ is almost twice less than Belle 
result. One can attribute the difference to any source of uncertainty 
described above. But another source the disagreement can arise from the 
higher fock state of $\chi_{c0}$ meson. Really it is known from NRQCD that 
color octet contribution of $\chi_{c0}$ meson is of the same order in 
relative velocity expansion as color singlet. Moreover it is known 
that color octet state gives NLO contribution to the amplitude, i.e. 
it changes main contribution to the cross section. So it would be 
interesting to estimate the size of this correction. 

\section{Discussion.}

In this paper the calculation of the cross sections of the processes $\epem\to J/ \Psi \chi_{c0}$ and 
$\epem\to \Psi(2S) \chi_{c0}$ at energy $\sqrt s = 10.6$GeV in the framework of light cone formalism has 
been carried out. It is shown that regarding the internal motion of mesons in the hard part 
of the amplitude in the framework of light cone results to the considerable enhancement of the cross 
sections in comparison to the NRQCD where internal motion is disregarded. So NRQCD is unreliable for 
the calculation of these cross sections. 

In addition to the processes $\epem\to J/ \Psi \chi_{c0}, \Psi(2S) \chi_{c0}$ Belle and BaBar experiments 
have measured the cross sections $\epem\to J/ \Psi \eta_c, \Psi(2S) \eta_c, J/ \Psi \eta_c(2S), 
\Psi(2S) \eta_c(2S)$. In the framework of light cone formalism these reactions were considered
in paper \cite{Braguta:2005kr}. The results obtained in this paper are presented in Table 2. 
Comparing the results for the cross sections measured in Belle and BaBar with light cone and NRQCD
predictions one can claim that despite a number of uncertainties the results obtained in the framework 
of light cone are in better agreement with Belle and Babar results than NRQCD predictions.

\begin{table}
$$\begin{array}{|c|c|c|c|c|}
\hline
 H_1 H_2 & \sigma_{BaBar} \times Br_{H_2 \to charged >2}( \mbox{fb} )  \mbox {\cite{Aubert:2005tj}}
 & \sigma_{Belle}\times Br_{H_2 \to charged >2}(\mbox{fb}) \mbox {\cite{Abe:2004ww}} 
 & \sigma_{LO} (\mbox{fb}) \mbox{\cite{Braguta:2005kr}}  
 & \sigma_{NRQCD} (\mbox{fb}) \mbox{\cite{Braaten:2002fi}}\\
\hline
\JP(1S) \ec(1S)& 17.6 \pm 2.8^{+1.5}_{-2.1} & 25.6 \pm 2.8 \pm 3.4 & 26.7  &2.31\\
\hline
\JP(2S) \ec(1S)&  - & 16.3 \pm 4.6 \pm 3.9 & 16.3 &0.96\\
\hline
 \JP(1S) \ec(2S)& 16.4 \pm 3.7^{+2.4}_{-3.0} & 16.5 \pm 3.0 \pm 2.4 & 26.6 &0.96\\
\hline
 \JP(2S) \ec(2S)& - & 16.0 \pm 5.1 \pm 3.8 & 14.5 &0.40\\
\hline
\end{array}$$
\caption{The second and third column contain experimental result. The results of paper 
\cite{Braguta:2005kr} are presented in the forth column. The last column contains NRQCD results.}
\end{table}

The authors thank V.L. Chernyak for  important remarks. This work was partially
supported by Russian Foundation of Basic Research under grant 04-02-17530, Russian Education
Ministry grant E02-31-96, CRDF grant MO-011-0, Scientific School grant SS-1303.2003.2. One of
the authors (V.B.) was also supported by Dynasty foundation.

\newpage
\begin{figure}[ph]
\begin{picture}(150, 200)
\put(-50,-100){\epsfxsize=9cm \epsfbox{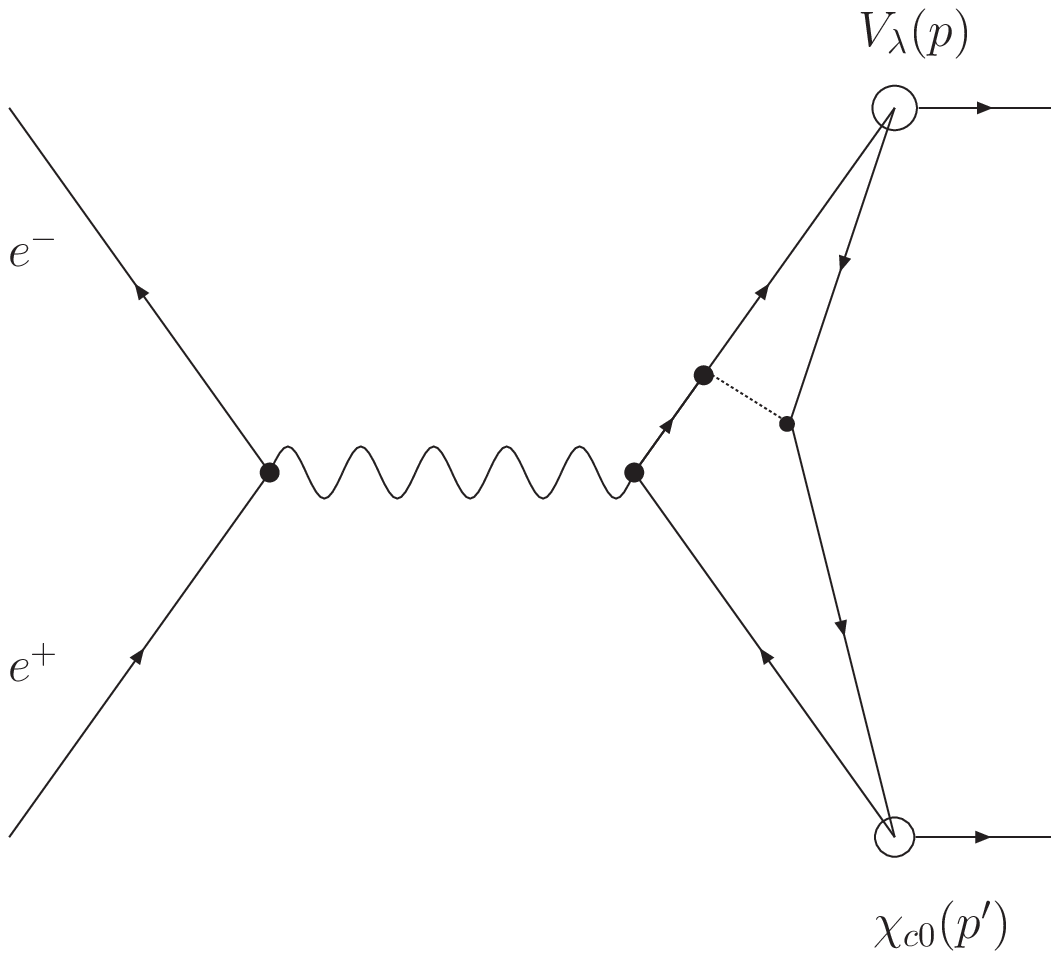}}
\put(-50,-350){\epsfxsize=9cm \epsfbox{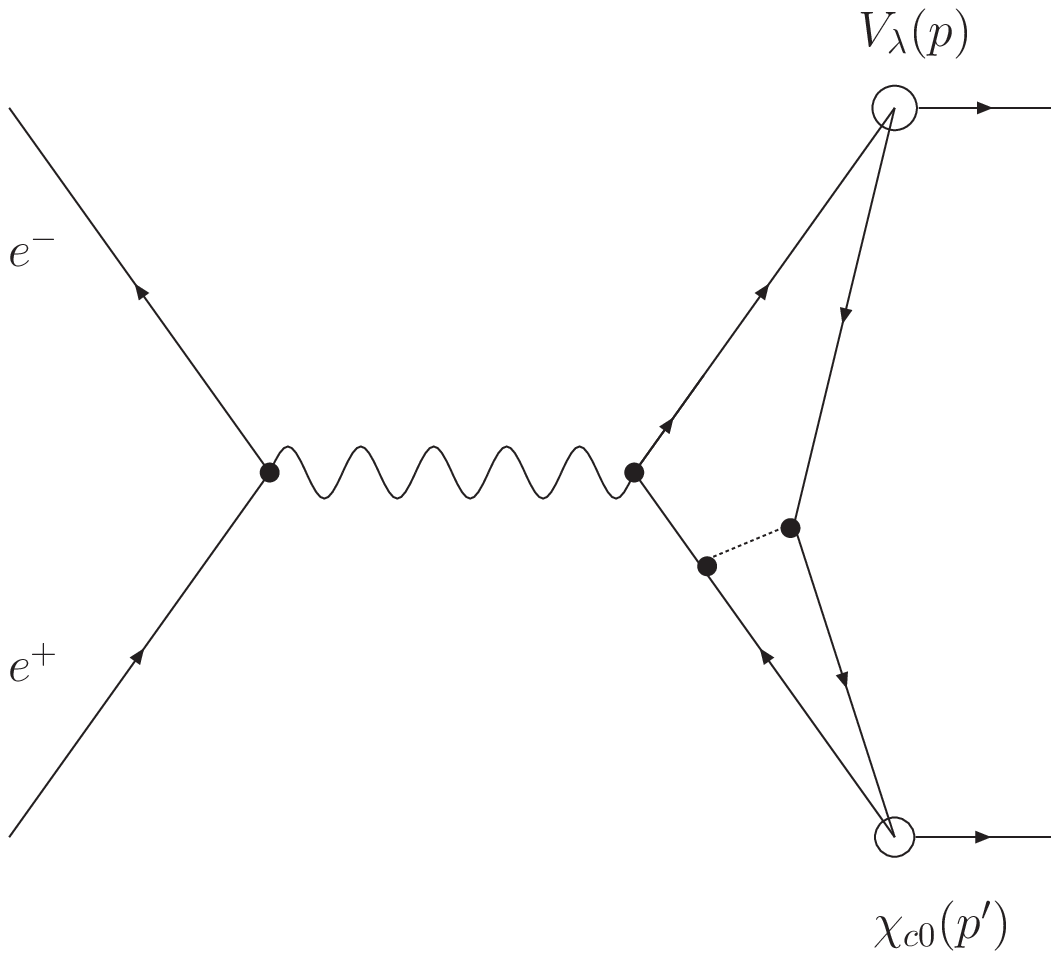}}
\put(-110,-370){\bf{Fig. 1: The diagrams that contrubute to the processes $\epem\to J/ \Psi \chi_{c0}, \Psi(2S) \chi_{c0}$}}
\end{picture}
\end{figure}

\end{document}